\input{aipcheck}
\documentclass[usenatbib,final,numberedheadings]{aipproc}
\usepackage{aas_macros}

\layoutstyle{8x11double}

\begin{document}

\title{Modeling Gamma-Ray Attenuation in High Redshift GeV Spectra}


\author{Rudy C. Gilmore}{
  address={University of California, Santa Cruz}
}

\author{Piero Madau}{
  address={University of California, Santa Cruz}
}

\author{Joel R. Primack}{
  address={University of California, Santa Cruz}
}

\author{Rachel S. Somerville}{
  address={Space Telescope Science Institute, Baltimore, MD} 
}

\begin{abstract}
 We present two models for the cosmological UV background light, and calculate the opacity of GeV gamma--rays out to redshift 9.  The contributors to the background include 2 possible quasar emissivities, and output from star--forming galaxies as determined by recent a semi--analytic model (SAM) of structure formation.  The SAM used in this work is based upon a hierarchical build-up of structure in a $\Lambda$CDM universe and is highly successful in reproducing a variety of observational parameters.  Above 1 Rydberg energy, ionizing radiation is subject to reprocessing by the IGM, which we treat using our radiative transfer code, CUBA.  The two models for quasar emissivity differing above z = 2.3 are chosen to match the ionization rates observed using flux decrement analysis and the higher values of the line-of-sight proximity effect.  We also investigate the possibility of a flat star formation rate density at z $>5$.  We conclude that observations of gamma--rays from 10 to 100 GeV by Fermi (GLAST) and the next generation of ground based experiments should confirm a strongly evolving opacity from $1<$ z $<4$.  Observation of attenuation in the spectra of gamma--ray bursts at higher redshift could constrain emission of UV radiation at these early times, either from a flat or increasing star-formation density or an unobserved population of sources.
\end{abstract}

\maketitle

\section{Introduction}

The Extragalactic Background Light (EBL) is the totality of light emitted by galaxies over the history of the universe.  High-energy gamma-rays propagating over cosmological distances are attenuated due to electron-positron pair production with EBL photons.  The optical depth tends to increase with gamma-ray energy, which modifies the high-energy spectra of distant sources.  The cross section for this annihilation process peaks at a wavelength of 1.24($E_\gamma$ /TeV) microns.  For Fermi (GLAST) energies (< 300 GeV), the relevant wavelengths are in the ultraviolet, and for energies below about 100 GeV the ionizing background is responsible for the pair-creation interaction.  A previous calculation of the background in the UV found that the contribution from resolved galaxies could make the universe optically thick to $\sim$30 GeV photons \cite{madau&phinney96}. 

In the optical and near-UV, the EBL is constrained by firm lower limits from number counts, e.g. \citep{madau00,gardner00}.  Indirect upper limits have recently been placed on the optical and near-IR background from ground-based gamma-ray experiments e.g. \citep{aharonian06,albert08}.  However, little is known about the background at energies above the Lyman limit from direct measurements.  Most ionizing photons from star-forming galaxies are absorbed by local cold gas and dust, with an uncertain fraction $f_{esc}$ escaping to the intergalactic medium.   Predicting optical depths to lower energy gamma-rays is further complicated by the fact that the attenuation edge for an optical depth of unity increases to a redshift of several, meaning that the evolution of ionizing sources must be understood to high redshift.  Beyond a redshift of about 2, the faint-end of the quasar luminosity function remains highly uncertain, with limits placed on the ionizing background via observations of the hydrogen and helium Lyman opacities.  The rapid increase of neutral helium above z$\sim$3 is believed to signal a strong decrease in the quasar emission of hard spectrum UV radiation.  While stars are thought to be the dominate source of UV photons at higher redshift, uncertainty in star-formation rates and efficiency, and possibility of changing escape fraction or initial mass function make predicting optical depths for gamma rays at high redshift much more difficult than studies of absorption with local TeV sources. 

\section{Modeling}

\begin{figure}
\resizebox{1.0\columnwidth}{!}{\includegraphics{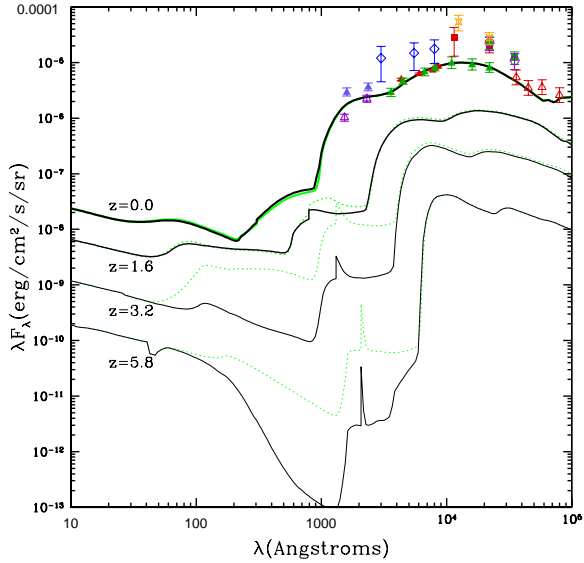}}
\caption{The background (from x-rays to IR) for the fiducial (black) and quasar-dominated (gray; green in online edition) models at the indicated redshifts.  The backgrounds at nonzero redshift are for photon populations evolved to present-day.  The data include lower limits from GALEX, HST, and IRAC (upward pointing arrows).  Direct detection data shown with other symbols include HST and DIRBE data with subtraction of zodiacal light and galactic sources.}
\label{figure:eblhist}
\end{figure}
\begin{figure}
\resizebox{0.9\columnwidth}{!}{\includegraphics{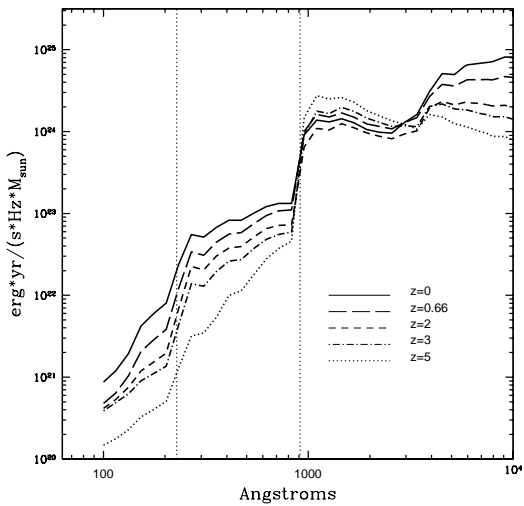}}
\caption{The total emissivity from star-forming galaxies in our semi-analytic model \cite{somerville08}, at the given redshifts, scaled to 1 $M_\odot yr^{-1}$.  HI and HeII ionization energies are indicated by thin vertical lines.}
\label{fig:starsed}
\end{figure}

We have calculated the EBL from the infrared to soft X-rays out to redshift 9.5 for 3 scenarios varying in escape fraction, quasar emissivity, and high redshift star formation rate.  Our approach utilizes recent semi-analytic modeling work \cite{somerville08} to determine stellar emissivities from galaxies.  These models are based on $\Lambda$CDM with h = 0.7, $\Omega_m$= 0.3, and use prescriptions based upon N-body and hydrodynamic simulations together with normalization from low-redshift data.  The models contain all the ingredients thought to be important in the evolution of galaxies.   A successful model of the UV background must take into account reprocessing of ionizing photons by the intergalactic medium (IGM).  Neutral hydrogen and neutral and singly-ionized helium in IGM clouds act as both sinks and sources of ionizing radiation \citep{haardt&madau96}.  We have used CUBA \citep{haardt&madau01} (an update to this software is in preparation), a numerical code for determining the propagation of UV photons through the IGM, to calculate the background as well as relevant observable parameters such as ionization rates and fractions.   The quasar luminosity evolution is based upon the 3 models presented in \citep{schirber&bullock03} (see Figure 4).  We assume a power law quasar spectrum with slope 1.57 shortward of 1200 Angstroms, as suggested by HST studies of quasars\citep{telfer02}.  We have developed 3 models for the evolution of the UV background that explore possibilities in star formation and quasar output; see Figures 1 through 4.  

For our fiducial model, we have used the observationally-based quasar evolution model of \citep{hopkins07}, with a moderate $f_{esc}$ of 0.1.  It should be pointed out that this model produces emissivities that are quite similar to those of model `A' in \cite{schirber&bullock03}.  In this case, stars and quasars produce enough ionizing photons to match ionization rate data from analysis of Lyman alpha flux decrement. 

The quasar--dominated model is based upon quasar model `C' in \cite{schirber&bullock03}, and is motivated by the higher ionization rates inferred from measurements of the line-of-sight proximity effect.  We also use a minimal $f_{esc}$ of 0.02 in this case.

A third possibility we have considered in our fiducial model is to modify the steeply declining comoving star-formation rate density in our semi-analytic model to remain flat for z$>$5.  

\begin{figure}
\resizebox{.9\columnwidth}{!}{\includegraphics{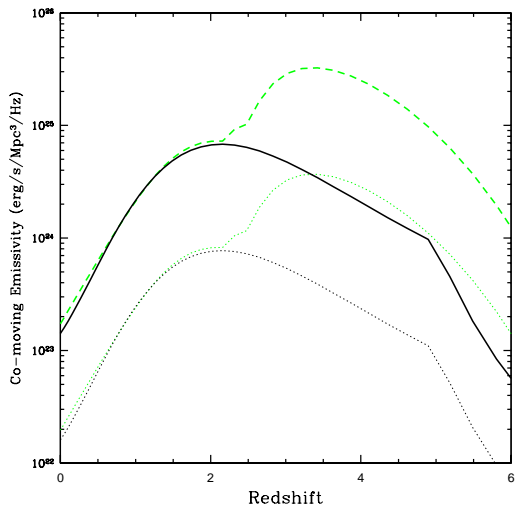}}
\caption{The cumulative emissivity density of quasars at 912 (upper) and 228 (lower) Angstroms in our models.  The black curve is our fiducial model, based on \citep{hopkins07}, and the dashed gray/green curve is based upon the model `C' in \citep{schirber&bullock03} multiplied by a factor 0.8. }
\label{fig:emiss}
\end{figure}
\begin{figure}
\resizebox{.9\columnwidth}{!}{\includegraphics{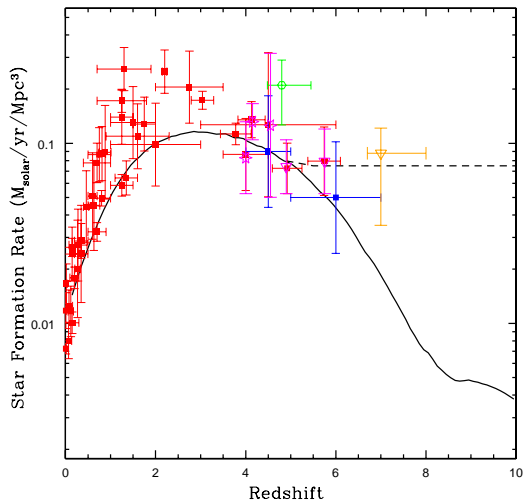}}
\caption{The star formation rate density in our semi-analytic model.  The dashed line showes the flat SFR density we consider as a possible modification to the declining SFR density of our model \cite{somerville08}.  Data is scaled to our Chabrier initial mass function.}
\label{fig:sfr}
\end{figure}

\section{Discussion and Conclusions}

\begin{figure}[t]
\resizebox{0.9\columnwidth}{!}{\includegraphics{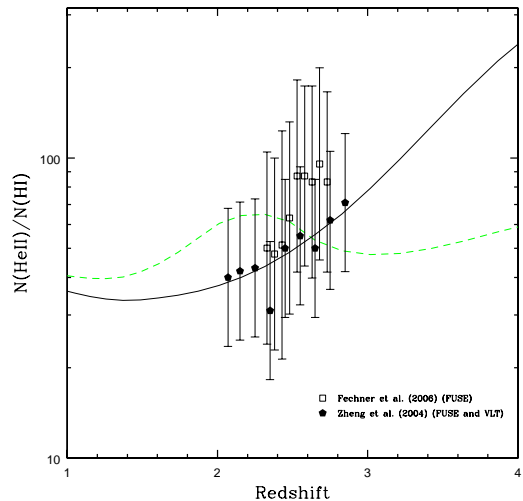}}
\caption{The ratio of column densities HeII/HI (softness) arising in each of our models.  The much more rapid rise for our fiducial model is indicative of its faster transition to a stellar-dominated ionizing background above z$\sim$2.5.}
\label{fig:softness}
\end{figure}

\begin{figure}[t]
\resizebox{1.0\columnwidth}{!}{\includegraphics{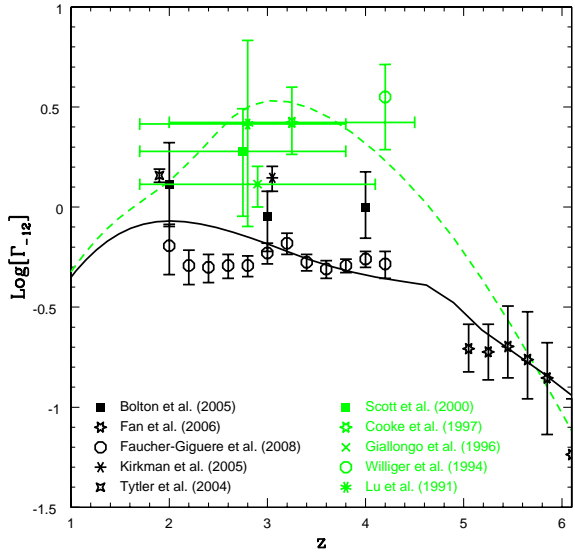}}
\caption{The ionization rate $\Gamma_{-12}$ of hydrogen in units of $10^{-12}$ $s^{-1}$, for each of our models.  As before, our fiducial model is in black, the quasar dominated model in dashed gray (green).  Data in black is from flux decrement measurements of the Lyman alpha forest.  The generally higher gray/green points are from measurements of the line-of-sight proximity effect, and motivated our consideration of the higher emissivities in the second model.}
\label{fig:ionrates}
\end{figure}

We have presented models of the UV background at all wavelengths relevant to absorption of gamma-rays in the Fermi LAT energy range (0.02 to 300 GeV).  Major uncertainties in UV emission include the poorly constrained escape fraction from star-forming galaxies and the faint end of the quasar luminosity function above $z \sim 2$, and we have explored the resulting EBL and and consequent gamma-ray absorption out to redshift 9 in three cosmological scenarios varying in these parameters.  Our fiducial model is based upon the quasar luminosity functions of \citep{hopkins07}.  While the faint end of the quasar luminosity function remains unobservable at these redshifts, it is argued in this work that bright quasars at the primary contributor to the total quasar emissivity in this epoch, and the contribution from unseen faint quasars is small.  The ionizing flux from quasars must also decrease rapidly above z$\sim$3 to account for sharply increasing helium Lyman-opacity, and must constitute only a small fraction of the total ionizing flux by z$\sim$6 \citep{srbinovsky&wyithe07}.  With a moderate contribution to the background from escaping radiation from star-forming galaxies, this model produces good agreement with UV softness measurements (Figure \ref{fig:softness}) and the low ionization rate (Figure \ref{fig:ionrates}) inferred from recent flux decrement measurements, cf. \cite{fg08}.   Our `quasar--dominated' model is based upon model  `C' in \citep{schirber&bullock03} and leads to the more absorption at low and intermediate redshifts.  Inclusion of this scenario is motivated by the ionization rate calculated from observations of the line-of-sight proximity effect, which are typically 2--3 times higher than those derived by other methods.  A large stellar ionizing component at this flux would violate observational limits on $f_{esc}$ \citep{siana07}.  A number of potential biases exist in this calculation which could account for the the higher $\Gamma_{-12}$, though the flux decrement method is not without biases either \cite{fg07}.  The quasar--dominated scenario also has trouble reproducing the trend of increasing UV softness approaching the redshift of helium reionization.   

\begin{figure}
\resizebox{1.0\columnwidth}{!}{\includegraphics{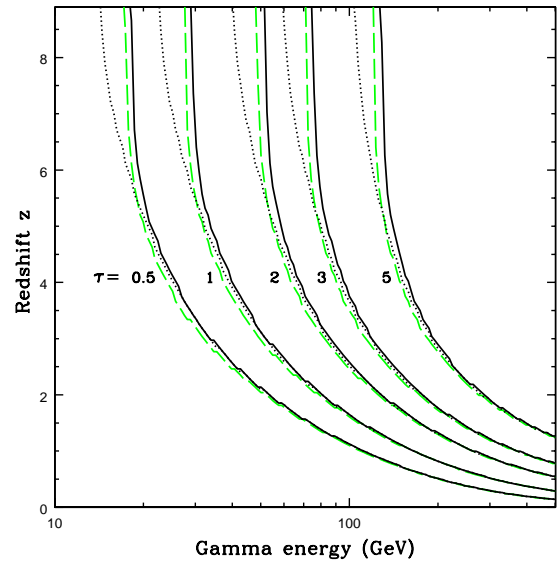}}
\caption{The attenuation edges for gamma--rays in each of our models.  The fiducial and quasar--dominated models are indicated using solid and dashed lines respectively, and the dotted lines show the model with a flat SFR density above $z=5$ (Figure \ref{fig:sfr}). Each set of curves indicates the redshift at which the optical depth $\tau$ reaches the indicated value (0.5, 1, 2, 3, and 5, from lower left to upper right).}
\label{fig:attedge}
\end{figure}

As an additional possibility, we have modified our fiducial model to have a flat star formation rate at $z>5$.  Predictions for gamma-ray absorption in these 3 scenarios are shown in Figure \ref{fig:attedge}.  All of our models create an optical depth that changes greatly between $z=1$ and 4.  The modified SFR density at $z>5$ leads to significantly greater optical depth at these redshifts, an effect that may be discernable if sufficient high-z VHE sources can be observed.  The mean free path of ionizing radiation is expected to fall quickly above z$\sim$6, as suggested by the appearence of complete Lyman alpha Gunn-Peterson troughs \citep{fan06}.  However, the $(1+z)$ multiplier from observed to rest-frame energy means that photons observed in the 10--100 GeV energy decade are affected primarily by non-ionizing UV and optical photons at these high redshifts.  This means that significant absorption of gamma-rays in this decade of interest can occur due to radiation below the Lyman limit even during reionization when the universe is largely opaque to ionizing UV photons.  While most data from the highest redshift Lyman-break galaxies has suggested a declining SFR \citep{bouwens08}, other techniques such as measurement of gamma-ray burst distributions \citep{yuksel08} have found higher rates, and little data exists at these redshifts.  An upcoming publication will examine different high-redshift scenarios in much greater detail, and in the context of reionization.

\begin{theacknowledgments}
We thank CUBA author Francesco Haardt for assistance in running the program and interpreting results, and Jeremiah Ostriker for useful discussions of reionization.
\end{theacknowledgments}

\end{document}